\documentclass[11pt,a4paper,fleqn,usenatbib,useAMS]{mnras}
\usepackage[T1]{fontenc}
\usepackage{ae,aecompl}
\usepackage{mathptmx}

\usepackage{graphicx}	
\usepackage{amsmath}	
\usepackage{amssymb}	
\usepackage{multicol}
\usepackage{pdflscape}	
\usepackage{placeins}
\usepackage{float}
\usepackage{xcolor}
\usepackage{enumitem}
\usepackage{ulem}
\usepackage{soul}
\usepackage{url}
\usepackage{breqn}

\title[Hot dust in NGC 1365 by Polar- RAT model]{Unravelling the  Nuclear Dust Morphology of NGC 1365 : A Two Phase Polar - RAT Model for the Ultraviolet to Infrared Spectral Energy Distribution }

\author[Subhashree Swain et al.]{
Subhashree Swain$^{1}$,
P. Shalima$^{2}$,
K.V.P. Latha$^{1}$\thanks{E-mail: lathakvp@gmail.com}
\\
$^{1}$Pondicherry University, Puducherry, 605014,India\\
$^{2}$Manipal Centre for Natural Sciences, Centre of Excellence, Manipal Academy of Higher Education, Manipal, Karnataka 576104, India\\
}

\pubyear{2022}

\begin{document}
\label{firstpage}
\pagerange{\pageref{firstpage}--\pageref{lastpage}}
\maketitle

\begin{abstract}

We present a 3D radiative transfer model for the Spectral Energy Distribution (SED) of NGC 1365, which is a ``changing look" Seyfert 1.8 type AGN. The SED from the ultraviolet (UV) to the infrared (IR) is constructed using archival data from the UVIT on-board \textit{AstroSat}, along with IR data from the literature. The {\sc skirt} radiative transfer code is used to model the SED and derive the geometry and composition of dust in this AGN. Similar to our earlier SED model of NGC 4151, the nuclear region of NGC 1365 is assumed to contain a ring or disk-like structure concentric to the accretion disk, composed of large (0.1$\mu$m - 1$\mu$m) graphite grains in addition to the two-phase dusty torus made up of ISM-type grains (Ring And Torus or RAT model). We also include, for the first time, an additional component of dusty wind in the form of a bipolar cone. We carry out a detailed analysis and derive the best-fit parameters from a $\chi^2 $ test to be $R_{\rm in,r}$ = 0.03 pc, $\sigma$ = 26$^\circ$ and $\tau$ = 20 for the assumed ring-torus-polar wind geometry. Our results suggest the presence of hot dust at a temperature T $\sim$ 1216 K at the location of the ring which absorbs and scatters the incident UV radiation and emits in the near-IR (NIR). In the mid-IR (MIR) the major contributors are the polar cone and warm dust with T $\sim$ 916 K at $R_{\rm in,t}$ = 0.1 pc. Not only are our model radii in agreement with IR interferometric observations, our study also reiterates the role of high resolution UV observations in  constraining the dust grain size distribution in the nuclear regions of AGN.
\end{abstract}



\section{Introduction} 
\label{sec:intro}
The active galactic nuclei (AGN) are one of the brightest extragalactic objects in the universe with luminosities $\backsim$ $10^{44}-10^{46}$ ergs\ s$^{-1}$ as 
they are powered by a supermassive blackhole (SMBH)($\backsim$\,10$^{6}$-10$^{9}$\,M$_\odot$). The AGN were classified as type 1 AGN where it shows both broad and narrow emission lines and type 2 AGN where it shows narrow lines only. The type 2 AGN mostly hides its continuum and broad emission lines with a surrounding dust toroidal structure called dusty torus, which is viewed in edge-on direction. The torus is thought to be located at a radial 
distance of $0.1-10$ pc \citep{doi:10.1146/annurev-astro-082214-122302} from the central blackhole. The radiation from the accretion disk mainly emits in 
the UV/optical regime (0.1$-$0.8 $\mu$m), which is absorbed and scattered by the dust grains in the torus. These heated dust grains then re-emit thermal radiation in the near-infrared (NIR) (1-5\, $\mu$m), mid-infrared (MIR) (5-25\, $\mu$m), and far-infrared (FIR) (25-500\, $\mu$m). This ultimately leads to the observed spectral energy distribution (SED) of AGN in the infrared (IR) region. 
The observations of several AGN in the IR band show a nearly flat/steep IR SED extending from 2$-$10 $\mu$m. Some AGN show NIR excess emission at 1-5 $\mu$\rm{m} \citep{1978ApJ...226..550R, 1987ApJ...320..537B, swain2003, suganuma2006, dexter2020} due to hot dust with silicate features at 10 $\mu m$ \citep{Deo2009} and an excess at $\sim$30 $\mu$m \citep{2011MNRAS.414.1082M}.  According to AGN unification schemes, the hot dust emission in the NIR cannot be directly observed in type 2 sources due to blocking of the radiation from the disk by the torus. However, it was reported that some type 2 AGN  also show NIR excess emission, which was then attributed to the radiation emitted by hot graphite dust in the sublimation region of the torus,  suggesting a clumpy torus \citep{0067-0049-204-2-23} rather than smooth distribution. Several authors \cite{selfcon, nenkova2008, 2017ApJ...838L..20H, Feltre2012, gracia2022, gomar2023} modeled the dusty torus considering smooth/clumpy/two-phase dust distribution to explain the IR observational features. However, it is a real challenge for modelling the SED of type 1.8/2 AGN with only a single geometry.\\
The nuclear region in active galaxies not only accretes matter onto the central SMBH but also contributes to starburst phenomena surrounding the AGN. The target AGN in this study, NGC 1365 is an AGN-starburst mix-type galaxy and a member of the Fornax cluster with a Cepheid-based distance of 18.6 Mpc \citep{Madore1999}. It has a variable obscured Seyfert 1.8 nucleus \citep{Risaltti2009}, a prominent bar and is an archetype barred spiral galaxy (SBb(s)I; \citep{1981RSA...C...0000S}). The nuclear region shows extended emission at different wavelengths. It is a highly absorbed source with a rapidly spinning black hole that exhibits significant relativistic disk reflection \citep{risaliti2013rapidly}. Extreme variability is observed on timescales of days and weeks, where it changed from Compton-thick to Compton-thin \citep{risaliti2007, Risaltti2009}.
In \cite{1999A&ARv...9..221L}, a comprehensive overview of NGC 1365 is provided. 

\cite{combesetal2019} have studied the large-scale CO(3-2) map of NGC 1365 and reported the existence of a central continuum point source, surrounded by a 
contrasted nuclear ring of radius $\sim$ 9\arcsec (= 770 pc). Inside this ring, the point source corresponds to a more compact molecular component, a circumnuclear disk of radius 0.3\arcsec(= 26$\pm$3 pc). This disk encircles the central continuum source and might be interpreted as the molecular torus of the AGN. \cite{Garbur2021} defined the nature of the extended components in NGC 1365 as only torus geometry using ALMA observations. According to \cite{herrero2021} this AGN has an unresolved polar dust component which was observed in the MIR using high angular resolution MIR and Atacama Large Millimeter/submillimeter Array (ALMA). This emission is dependent on the column density and Eddington ratio \citep{gracia2022}.  \\

 It is difficult to directly observe the inner radius of AGN NGC 1365 torus as the dust lane passing through the nuclear region. But \cite{tristam2011} succeeded in measuring the innermost radius of NGC 1365 AGN torus using the MID-infrared Interferometric instrument (MIDI) of the ESO Very Large Telescope Interferometer (VLTI) and determined the location of warm dust clouds in this AGN. 
 With IR interferometry \citep{swain2003, kishimoto2011} and reverberation mapping \citep{kishimoto2007, weigelt2012}, the innermost regions of some of the brightest type-1 AGN were partially resolved and their sizes were estimated. However, type 2 AGN is out of the list because of its high column density along the LOS towards the observer. However, our chosen target has a history of being a Seyfert 1, intermediate Seyfert and Seyfert 2 AGN. But, the study on the innermost radius of the sublimation zone was limited for this object. Thanks to VLTI/GRAVITY interferometer, which measured the compact sizes of a larger sample of AGN by NIR interferometry (\cite{dexter2020}) based on the luminosity-size relation \citep{suganuma2006, kishimoto2007}, as expected for the dust sublimation region. This team has resolved the nucleus in the brightest type-I AGN including NGC 1365 and found its NIR emitting region consisting of hot dust to be a ring of radius 0.03$\pm$0.004 pc. This agrees well with the values obtained for nearby type-I AGN from MIR reverberation mapping \citep{kishimoto2007, lyu2022}. Though their values were found to be smaller than the dust sublimation radius for silicate dust by a factor of 3, \cite{Kawaguchi_2011} showed that this discrepancy could be explained and corrected by considering the anisotropy of the accretion disc emission. Again, large graphite grains can survive much closer to the nucleus due to their higher sublimation temperatures compared to silicates, which can also account for the smaller observed radii. Hence, the lack of clear knowledge on sublimation radii for NGC 1365 demands an advanced model considering the effects of hot dust, polar dust, warm dust, etc. However, the dust sublimation zone in the nucleus of NGC 1365 is quite challenging to be probed, not only due to the lack of availability of high resolution instruments, but also because, this variable AGN is itself hosted in a starburst spiral galaxy. As a result, a variety of physical phenomena such as starburst activity, polar wind activity, and central AGN activity all together contribute to the SED. The 9.7 $\mu$m silicate feature in this AGN has been reported to be in slight emission, absorption \citep{martinetal2013} or flat and featureless \citep{herrero2020}, which are all within the error bars. Different levels of host galaxy contamination in the measurements has been suggested to be causing these differences \citep{martinetal2013}. It is practically difficult to disentangle all the possible contributions to the central AGN luminosity. Hence, it is crucial to investigate the emission from the innermost regions of this AGN using high resolution observations at other complementary wavelengths like the UV, for better and more realistic radiative transfer modelling and SED fitting. \\
It is known that the torus emits in IR by absorbing UV photons from the accretion disk \citep{Netzer2015, almeida2011testing}.
There have been several attempts in the past to model the SED of AGN torus and infer its geometry and dust composition. \citet{Ramos_Almeida_2011} have derived the geometry and dust distribution of a sample of AGN including NGC 1365 by using the clumpy torus model from which, parameters like half opening angle of the torus ($\sigma$), inclination angle (\textit{i} ), outer radius (R$_{\rm out}$), optical depth ($\tau_v$), etc. were obtained. 
This was followed by \cite{hererroetal2012, selfcon, 2017ApJ...838L..20H} who considered clumpy, multi-phase polar models respectively to reproduce the observed SED. \cite{subhashree2021} has used the two-phase torus model using SKIRT code \citep{stalevski2012} with an additional graphite ring-like structure close to the accretion disk which accounted for the NIR emission in NGC 4151. Similarly \cite{gracia2022} have considered the CAT3D-Wind model of \cite{2017ApJ...838L..20H} with an additional component of hot large graphite grains blended with the MIR emitting region close to the nucleus to fit the IR SED of NGC 1365. More recently, \cite{gomar2023} found the grain size distribution to be another important parameter after half opening angle and torus geometry by constructing a library of models to fit the IR SED of a large sample of AGN.  An additional ring geometry has also been found necessary to explain the AGN SED in some cases \citep{subhashree2021}. However, the two-phase model is a more convincing picture of dust around luminous AGN since both smooth and clumpy dust grains help in shielding the smaller grains to survive along with large grains. Hence it is one of the more realistic models in the case study on the dusty torus where the different grain size distributions can explain the NIR and MIR features including the silicate features. Hence, in the present work, different dust compositions of different size distributions were explored separately for three geometries around the disk i.e. graphite ring, torus and polar cone. We focus on the grain size distribution with this modified geometry for NGC 1365 with supporting high resolution UV observations.
In the present study, we have constructed and modeled the UV to IR SED of NGC 1365 for the first time, by including recent FUV and NUV observations from the Ultra-Violet Imaging Telescope (UVIT), a multi-band instrument of India's {\textit{AstroSat}} mission. The high spatial resolution of UVIT presents the unique advantage of extracting the UV flux of the AGN with minimum contamination from the surrounding star-forming regions. The IR data is taken from Spitzer MIPS for 24 $\mu$\rm{m} and {\sc Hershel}/PACS imaging data for 70 $\mu$\rm{m}, 160 $\mu$\rm{m}, 250 $\mu$\rm{m}. We follow the method outlined in \citet{subhashree2021} for modeling the UV-IR SED of this AGN using the \cite{stalevski2016} model with a graphite ring for the NIR emission, and an additional polar wind component for the MIR emission.

\section{Data and observations}
 The \textit{AstroSat} 
 \citep{AGRAWAL20062989, singh2014} 
   carries four co-aligned instruments, Ultra-Violet Imaging Telescope (UVIT), Soft X-ray Telescope (SXT), Large Area X-ray Proportional Counter (LAXPC) and Cadmium-Zinc-Telluride Imager (CZTI). Archival UV data from UVIT and archival IR data from HST/NICMOS and GEMINI instruments are used in this work.

\subsection{UV data}
The observation log of UV data used here is given in Table \ref{tab:tab1}. 
\begin{table}
	\centering
	\caption{Observation log of the data used in our spectral modeling.}
	\label{tab:tab1}
	\begin{tabular}{lcccr} 
		\hline
		Observatory & OBSID & Exposure time \\
		\hline
		\textit{AstroSat}/UVIT/FUV & 9000000776 & 7.9 ks \\
		\textit{AstroSat}/UVIT/FUV & 9000000802 & 12.1 ks\\
		\textit{AstroSat}/UVIT/FUV & 9000000934 & 6.7 ks\\
		\hline
		\textit{AstroSat}/UVIT/NUV & 9000000776 & 9.9 ks \\
		\textit{AstroSat}/UVIT/NUV & 9000000802 & 11.8 ks\\
		\textit{AstroSat}/UVIT/NUV & 9000000934 & 6.4 ks\\
		\hline
	\end{tabular}
\end{table}
We used the deep NUV and FUV imaging observations of NGC 1365 from UVIT (\cite{Tandon2017}; \cite{Tandon2020}). The UVIT consists of two co-aligned telescopes, one for FUV (1300 \AA $-$ 1800 \AA) and
another for both NUV (2000 \AA $-$ 3000 \AA) and visible (VIS) channels (3200 \AA $-$ 5500 \AA). Each of
the pointing telescopes has a 28$\arcmin$ circular field of view with a high
angular resolution (FWHM 1$\arcsec - $1.5$\arcsec$).
NGC 1365 was imaged in FUV/F169M ($\lambda_{mean}$ = 1608 \AA, $\bigtriangleup\lambda$ = 290 \AA) and NUV/N279N ($\lambda_{mean}$ = 2792 \AA, $\bigtriangleup\lambda$ = 90 \AA) filters. 
\par
\begin{figure}
	\includegraphics[width=\columnwidth]{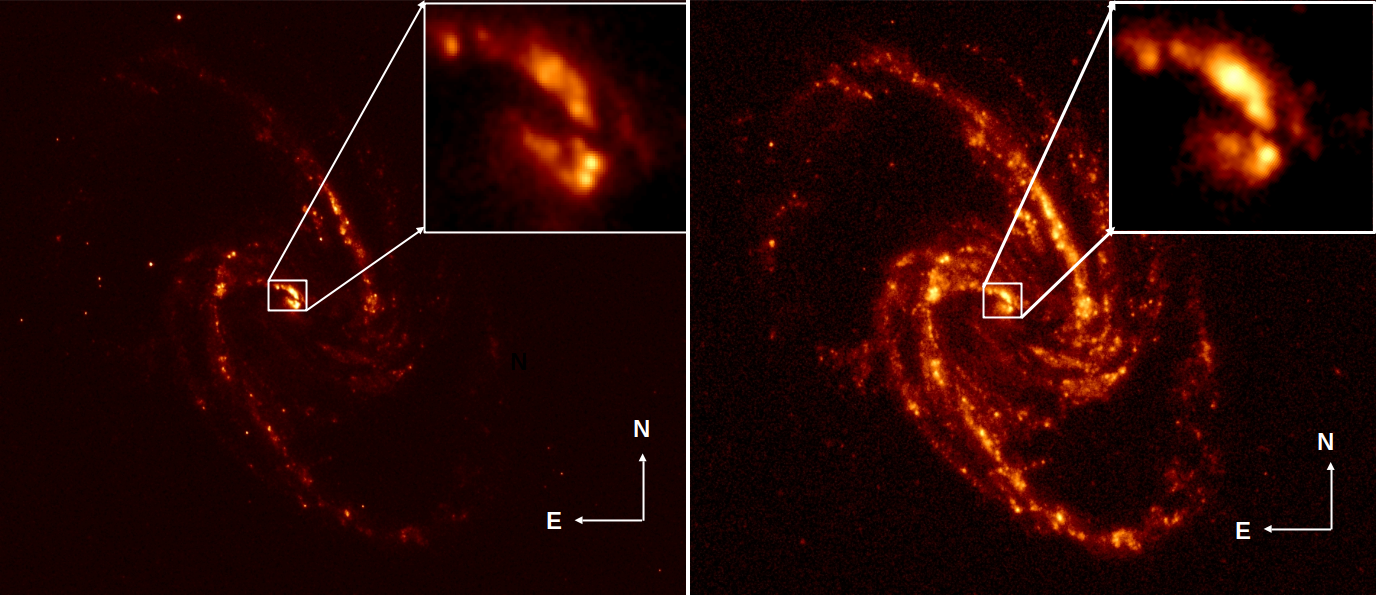}
    \caption{Left panel: Astrometry corrected NUV image with size 8\arcmin $\times$ 11\arcmin, Right panel: The zoomed central region with size 40$\arcsec$ $\times$ 45$\arcsec$, Right panel: Astrometry corrected FUV image with size 8\arcmin $\times$ 11\arcmin, Right panel: The zoomed central region with size 40$\arcsec$ $\times$ 45$\arcsec$. The central region in both UV bands is obscured by a dust lane. This is for the central view of NGC 1365. }
    \label{fig:fig1}
\end{figure}

The deep combined images of three observations in NUV and FUV bands shown in Figure~\ref{fig:fig1} are taken from \cite{swain2022}. The image sizes are 4096 $\times$ 4096 pixels with a pixel scale
of 0.4$\arcsec$ which corresponds to $\sim$ 36 pc at the distance of NGC 1365. The final images in both bands suggest that the central region of the AGN in the FUV is more extinct than that in the NUV band.
 The point source at the center of NGC 1365 is not visible in both the images due to the obscuration by the foreground dust. 
Therefore we followed the host galaxy subtraction method as in \cite{swain2022} where authors performed the 6" aperture photometry by surface brightness profile (SBP) method using Moffat function and then corrected it for galactic and internal extinction. Here, we calculated the intrinsic AGN count rate which was used further for the AGN SED. The UV emission did not show any variability during the observation period \citep{swain2022}. Although the accretion rate is very low, the FUV emission from the accretion disk could still be responsible for the observed dust emission.

\subsection{IR data}
The NIR data are taken from \cite{almeida_2009} where they used high spatial sub-arcsecond resolution observations from HST/NICMOS in H, K, L and M bands. These bands are useful in studying the hot dust grain components in AGN. The AGN flux in these bands are not completely free from  stellar emission. Although  3$\arcsec$ aperture size is considered in order to minimize the contribution from the host galaxy, stellar emission still dominates at wavelengths below 2.2 $\mu$\rm m. 
The HST/NICMOS data for AGN \citep{Alonso_Herrero_2001} reveals that L and M bands are dominated by non-stellar emission with lower stellar  contributions of 25\% and 10\% respectively, while H and K bands have  up to 50\% contributions from stellar emission.
In addition, we use MIR data from \cite{hererroetal2012} where the authors have carried out the analysis with T-ReCs/GEMINI instrument. To derive the SED spanning NIR to MIR energies, they have utilised data from Spitzer, Herschel and T-ReCs/GEMINI instruments. 
They have done the flux calibration and PSF subtraction to get the AGN emission. There is no clear indication of the 9.7 $\mu$\rm{m}~feature for this particular AGN from observations. There could be possible contamination from extended components (non-AGN emission) in the extraction aperture of 3.7"$\times$3.7" taken by \cite{hererroetal2012}.
\begin{table}
	\centering
	\caption{The Infra-red instrument details used in our study.}
	\label{tab:tab2}
\begin{tabular}{cccc}
\hline 
Instrument & $\lambda_c$ & Pixel size & Flux \\ 
 & ($\mu m$) & (arcsec) & (mjy) \\ 
\hline
NICMOS & 1.6 & 0.18\arcsec & 8.3$\pm$0.8 \\
\hline
NICMOS & 2.2 & 0.18\arcsec & 78$\pm$8 \\
\hline
NICMOS & 3.5 & 0.18\arcsec & 205$\pm$41 \\
\hline
NICMOS & 4.8 & 0.18\arcsec & 177$\pm$35 \\
\hline
T-ReCS & 8.7 & 0.35 & 203$\pm$30 \\
\hline
T-ReCS & 13.0 & 0.35 & 400$\pm$60 \\
\hline
T-ReCS & 18.3 & 0.35 & 818$\pm$205 \\
\hline
\end{tabular}
\end{table} 
\section{Methodology}
\begin{figure*}
	\includegraphics[width=16cm]{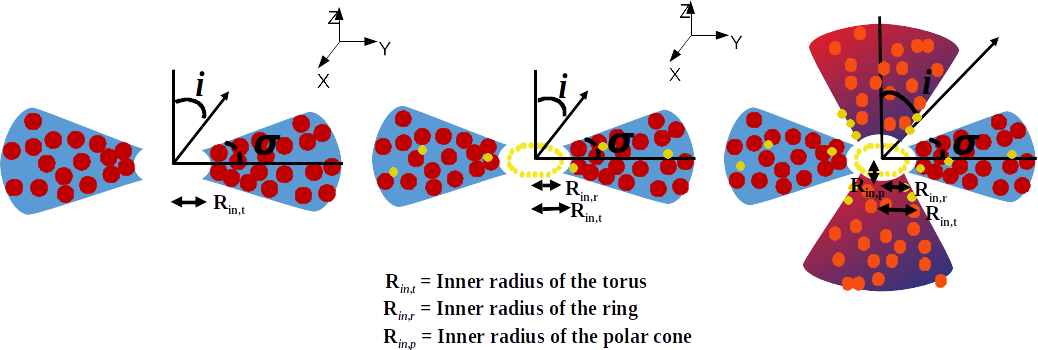}
    \caption{The Schematic diagram of the cross-sectional view of the TO, RAT and polar model
with `$\sigma$' is the half-opening angle and `i' is the inclination angle.}
    \label{fig:fig4a}
\end{figure*}

We have adopted the Torus model, Ring and Torus model, and Polar wind model for this work. Full description of the models
can be found in the \cite{subhashree2021}. These models are explained briefly in this section and illustrated in Figure \ref{fig:fig4a}.\\
\begin{itemize}
\item {\it Torus only model (TO)}: This model consists of a flared torus surrounding the central nucleus with the anisotropic disk. The torus geometry is confined with the inner radius, outer radius, half opening angle, number of clouds, the size of clumps and optical depth. The dust medium in the torus is two phase (smooth+clumpy), which consists of MRN mixture of ISM grains of size 0.005-0.25 $\mu$\rm m  with sublimation temperature of 1500 K. The two phase medium consists of a large number of high density clumps embedded in a smooth low density medium. \\
\item {\it Ring and Torus (RAT) model:} This model incorporates hot graphite dust in an additional ring like structure surrounding the accretion disk followed by the dusty torus. The ring consists of only large graphite particles in the region with sizes ranging from 0.1 to 1 $\mu$\rm m as pure graphite grains have a higher sublimation temperature of 1800 K and only large grains can survive in the innermost region of the torus \cite{laor_2018}. The ring is defined by its width, height and inner radius of the ring. In the current model, the torus geometry is detached from the ring geometry and hence inner radius of the torus does not coincide with the outer radius of the ring.
The models are named as smooth/clumpy RAT depending on whether the dust distribution inside the ring is smooth/clumpy respectively.
The geometry of the two phase torus is retained in these models. \\
\item {\it Polar wind model:} In addition to the graphite
ring, a conical shell along the polar axis has been incorporated in
order to include a polar wind, which is known to be responsible
for MIR emission \citep{herrero2021}. This is the new geometry introduced in this work. Here the model includes a polar cone with the RAT model. The polar cone is defined by its inner radius, outer radius, and the angle between the polar axis and the edge of the cone. The dust distribution inside the conical shell consists of silicate grains with sizes ranging from 0.04 $\mu$\rm m to 10 $\mu$\rm m \citep{Lyu2018}. 
\end{itemize}

\section{Results}
\subsection{UV-IR morphology}
\begin{figure}
	\includegraphics[width=4.2cm]{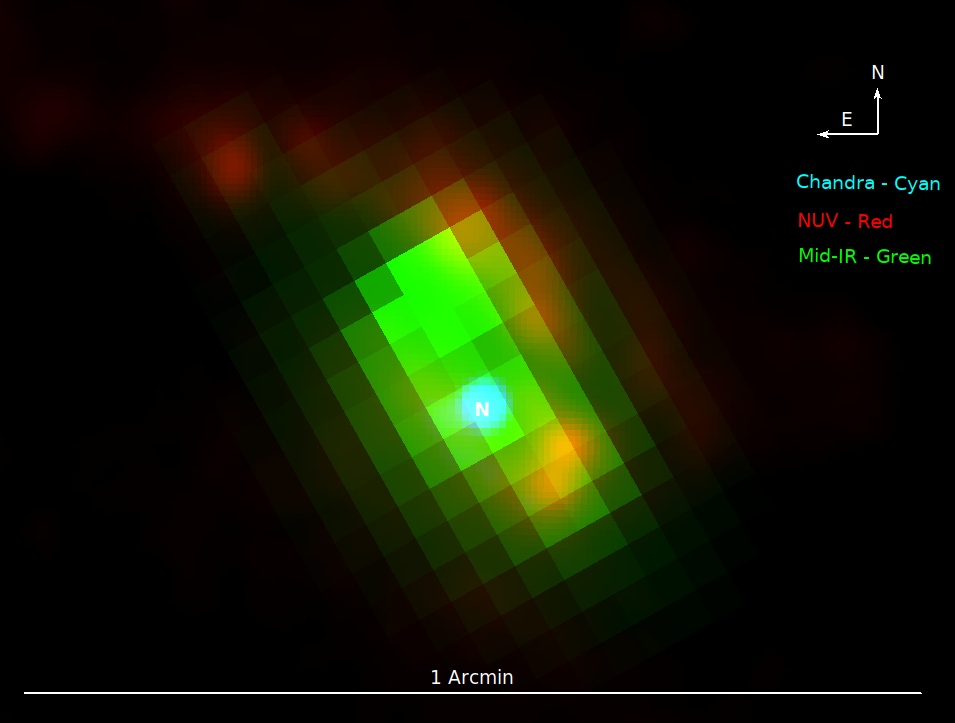}
        \includegraphics[width=4.2cm]{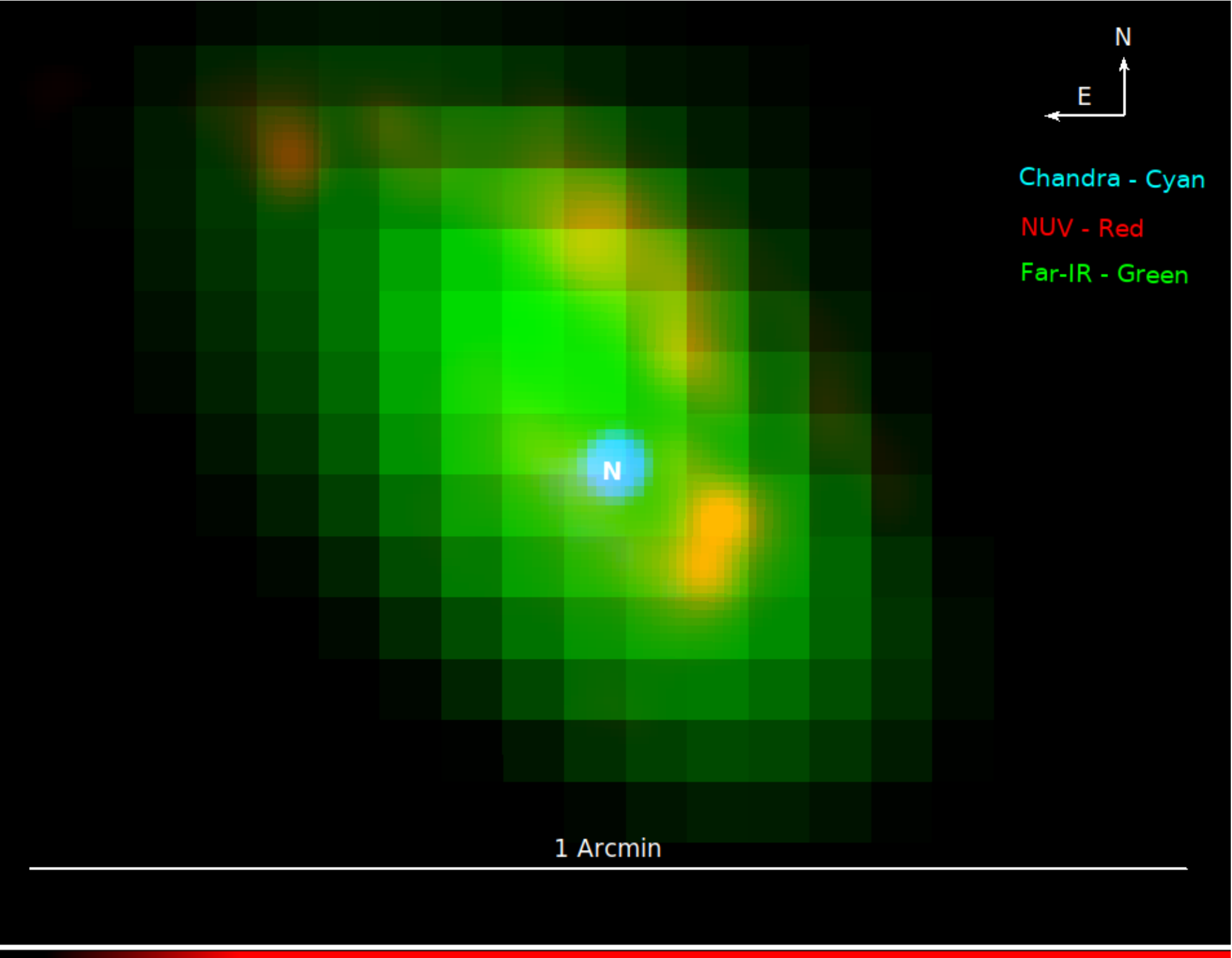}
    \caption{Left panel: The stacked image in NUV (red) and MIPS 24$\mu$m (green) of the central AGN. The `N' symbol represents the Chandra X-ray nucleus. Right panel: The stacked image in NUV (red) and PACS 160$\mu$m (green) of the central AGN. The `N' symbol represents the Chandra X-ray nucleus.}
    \label{fig:fig4}
\end{figure}

The images in the NUV, FUV, MIR (MIPS 24 $\mu$m), as well as FIR (PACS 160 $\mu$m) bands, are used to construct composite images of NGC 1365. The overlaid images of NUV and MIR bands are shown in the left panel of Figure \ref{fig:fig4} while that of NUV and FIR bands are shown in the right panel. The position of $Chandra$ nucleus is used to locate the central AGN as it is a clear point source detected by the instrument. The AGN is still bright in the MIPS 24 $\mu$\rm{m} band. Figure \ref{fig:fig4} shows a large extended bright region in PACS 160 $\mu$m band containing two bright NUV spots in the SW direction. FIR emission is extended up to a radius of 25$\arcsec$ ($\sim$ 2 kpc) from the center. 
\subsection{Modelling the UV-IR SED}
The multi-wavelength AGN SED is important for understanding torus geometry, composition, and  morphology, which are major components of AGN unification models. Therefore, we use
pure intrinsic flux from UVIT observations from \cite{swain2022}. The AGN contribution in the NUV band is weak in the core region within $\sim$ 6\arcsec aperture, while the FUV band is completely obscured by the dust lane passing in front of the nucleus. By incorporating the intrinsic UV flux in the AGN SED along with the IR observed flux, we can examine the disk-torus connection. In order to investigate the location of hot dust in NGC 1365 AGN and to construct its SED in the UV-IR wavelength range, the `Ring and Torus model' (RAT model) of \cite{subhashree2021} is used. There, the authors used the {\sc skirt} code \citep{stalevski2016,steinacker2013three, doi:10.1046/j.1365-8711.2003.06770.x, 0067-0049-196-2-22} that considers a two-phase smooth and clumpy medium to fit the NIR SED of NGC 4151. In the RAT model, two sublimation radii are taken into account corresponding to graphite and ISM-type dust respectively, with different size distributions instead of one sublimation radius. As a result, the pure graphite ring has a radius ranging from 0.02 pc to 0.06 pc pertaining to the large grain size range. Beyond the ring, is the torus that has the inner radius of the torus defined by the sublimation radius for the ISM type grains ranging from 0.03 pc to 0.2 pc. Depending on the grain size in a conical shell, the minimum radius of the polar cone ranges from 0.07 pc to 0.2 pc from the center.
As the geometry consists of grains of different sizes and temperatures, the inner radius of the ring is varied from 0.02 pc to 0.8 pc. We have extended the inner radius of the ring and torus beyond its theoretical value as the dust is always being pushed by central AGN.
The outer radius, R$_{\rm out}$ of the torus is assumed to be 5 pc. The ratio of outer radius of the torus to the inner radius of the dust sublimation zone is $\gamma$ $\sim$160, which is close to the values assumed by \cite{selfcon} ($\gamma$ = 150) and \cite{refId01} ($\gamma$ = 170). The total amount of dust is determined by the equatorial optical depth at 9.7$\mu$\rm{m}  which is used as an input for generating the clumps inside the geometry. For the
present model, the dust distribution parameters are fixed at p = 1 and q = 0 \citep{stalevski2016}. The parameter N$_{\rm clump}$ is the number of clumps inside
the torus, ring and polar cone where grains are present in a two-phase/clumpy/smooth medium, and F$_{\rm clump}$ is the fraction of total dust mass in the torus which is fixed at 0.97. The number and size of the clumps are usually chosen to
achieve a certain volume filling factor of the torus, ring and polar cone. The number of clumps has been fixed at $\sim$ 8000 in the torus and 4000 in the polar cone and varied in the ring. As a result, they could overlap with
each other and form high density complex structures. The
contrast parameter defined as the ratio of the high and low
densities of the medium, is kept constant at 100. The parameters varied to get the best-fit SED are given in Table \ref{tr1}.

\begin{figure}
\centering
	\includegraphics[width=8cm]{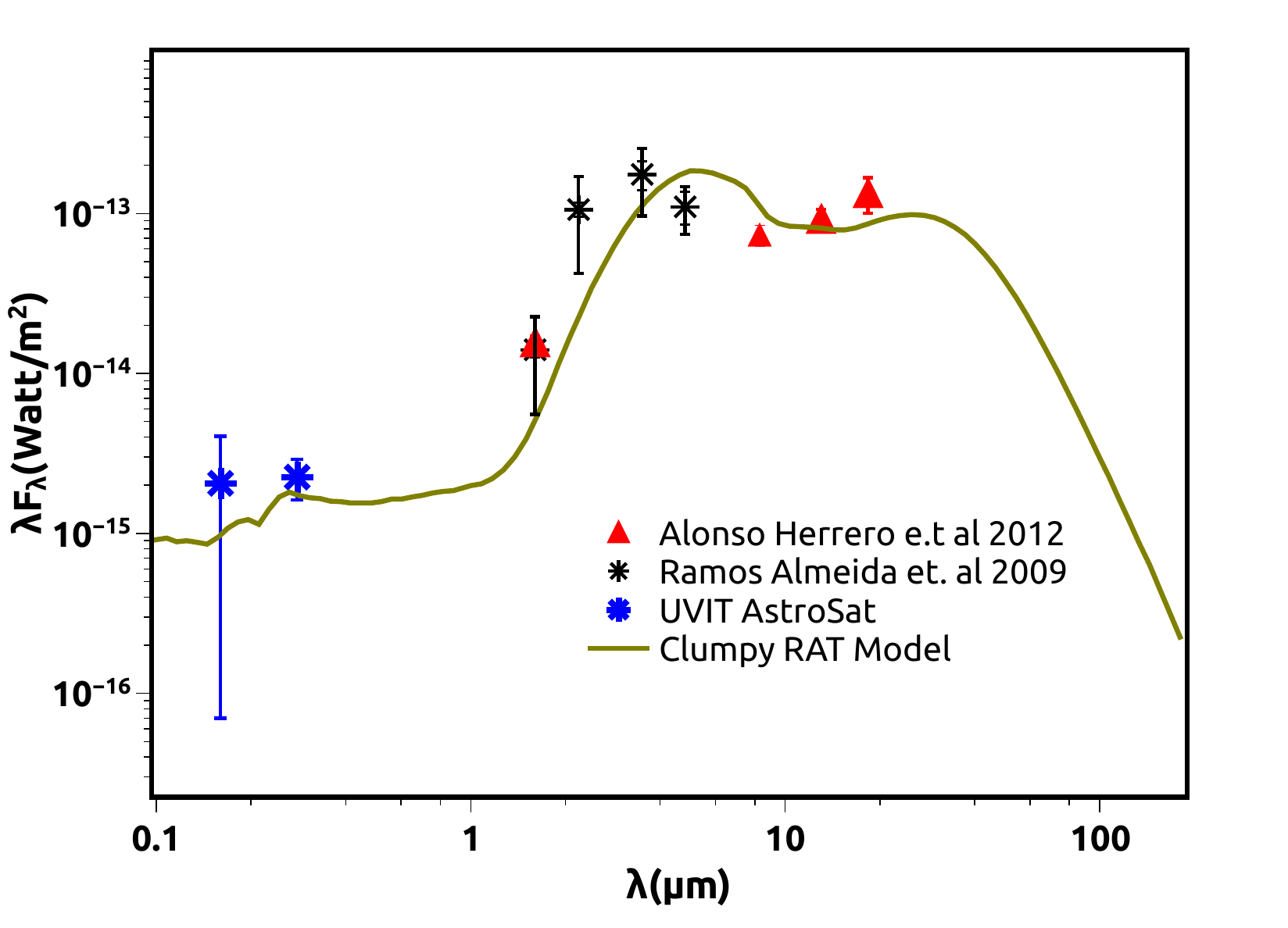} 	
\caption{Clumpy RAT model SED with best fit parameters at $\sigma$ = 26$^{\circ}$, $\tau_{9.7, \rm r}$=10, $\tau_{9.7, \rm t}$=10, $R_{\rm in, \rm r}$ = 0.4 pc, $R_{\rm in, \rm t}$=0.5pc, $i$= 69$^{\circ}$ and N$_{\rm clumps}$=3000.}
\label{fig:fig166}
\end{figure}

\begin{figure}
\centering
	\includegraphics[width=8cm]{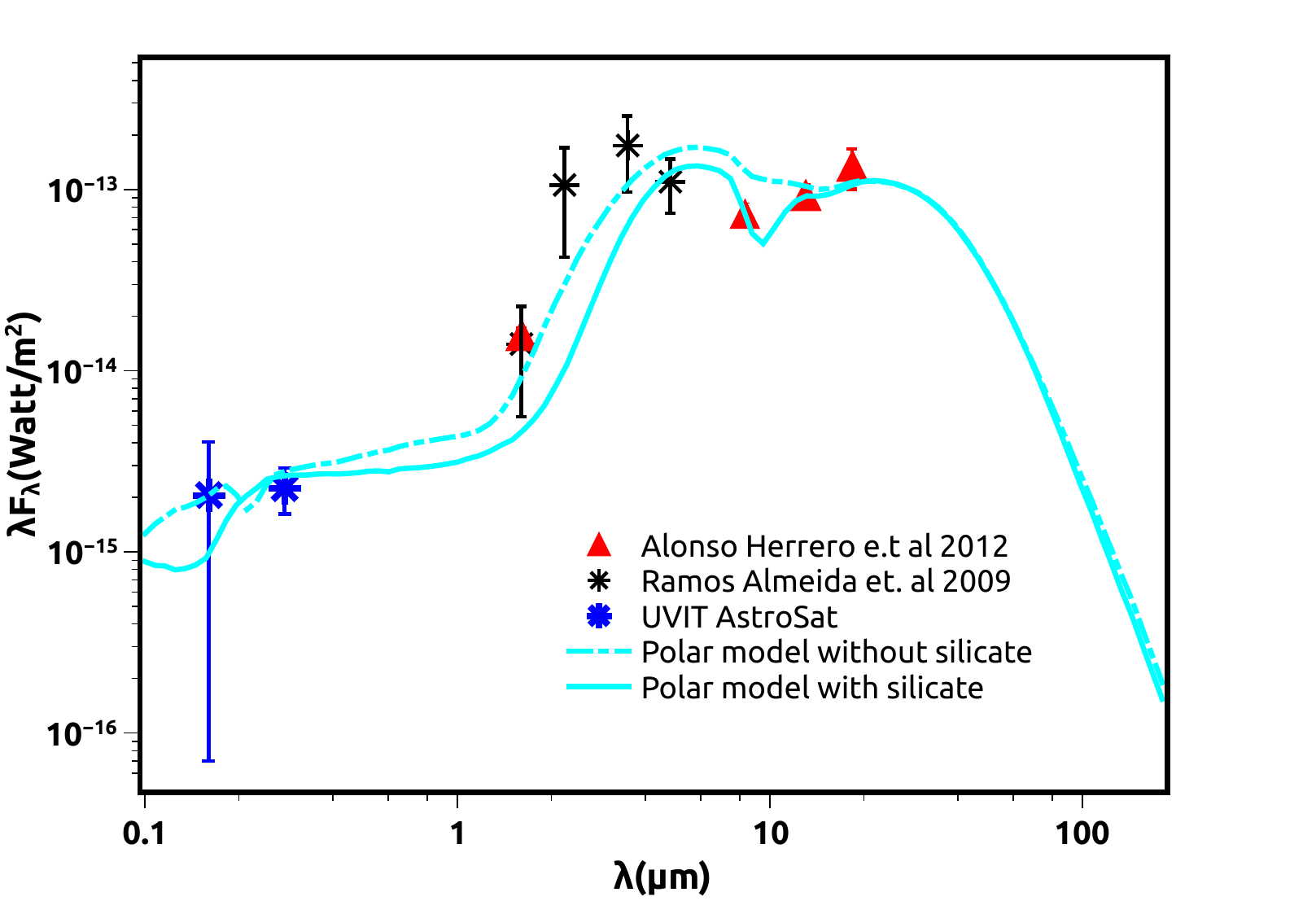} 	
\caption{Polar model SED with best fit parameters at
$\sigma_p$ = 30$^{\circ}$, $R_{\rm in, \rm r}$ = 0.03 pc, $R_{\rm in, \rm t}$ = 0.1 pc, $R_{\rm in, \rm p}$ = 0.08 pc, $i$ = 72$^{\circ}$.}
\label{fig:fig67}
\end{figure}

\begin{figure}
\centering
\includegraphics[width=8cm]{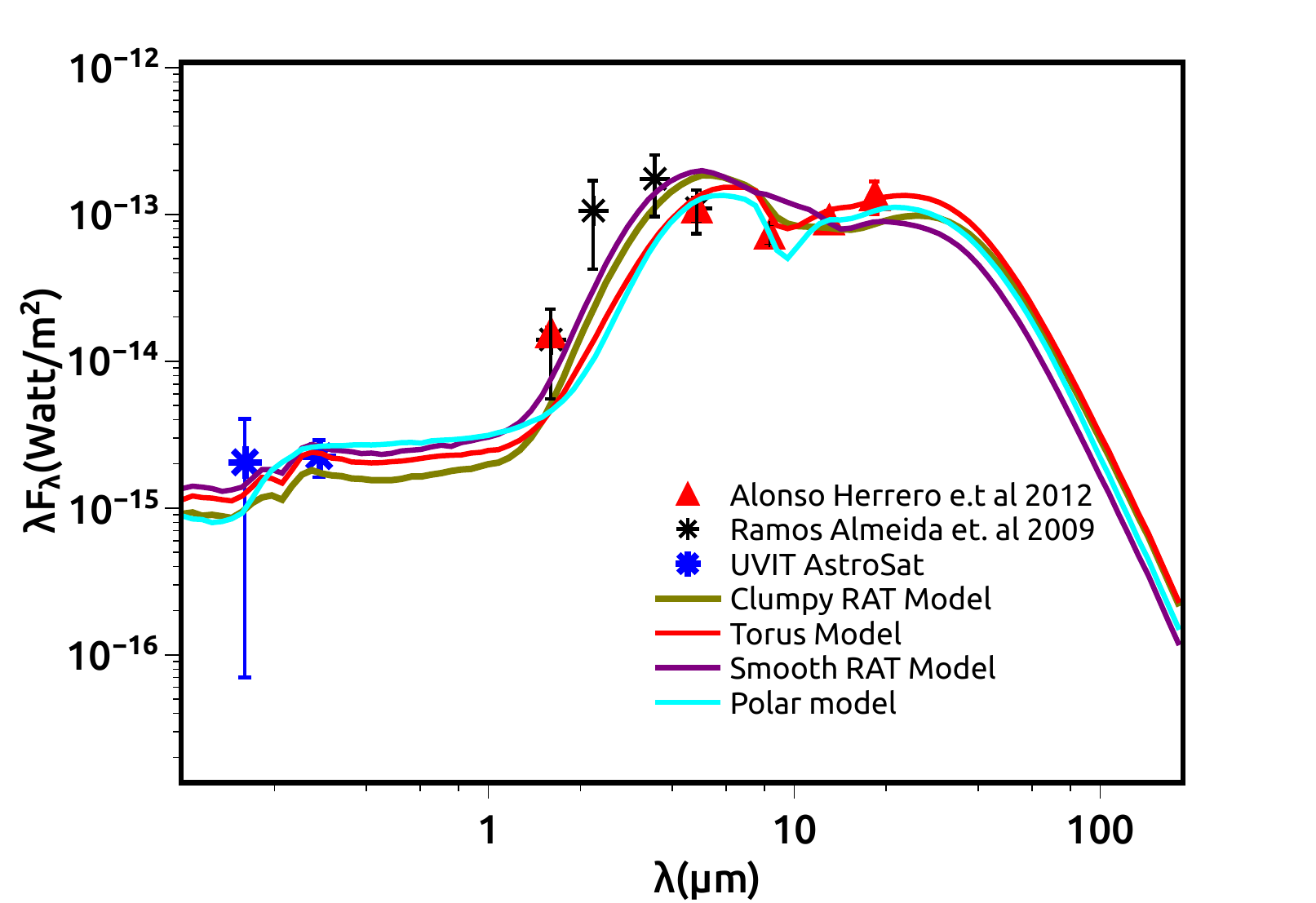}
\caption{The AGN SED with best fit parameters at
$R_{\rm in,r}$ = 0.4 pc, $\tau_{9.7, \rm r}$= 10, $\tau_{9.7, \rm t}$ = 10, $\sigma_t$=26$^{\circ}$, i=69$^{\circ}$ and N$_{clumps}$=3000 for smooth and clumpy RAT model, $R_{\rm in,t}$ = 0.4 pc, $\tau_{9.7, \rm t}$= 10 and $\sigma_t$=26$^{\circ}$ for Torus model and $R_{\rm in,r}$ = 0.03 pc, $\tau_{9.7, \rm t}$= 10 and $\sigma_p$=30$^{\circ}$, i=72$^{\circ}$ for polar model.}
\label{fig:fig511}
\end{figure}

\begin{table}
\begin{center}
\caption{Parameters used in the model}
\label{tr1}
\begin{tabular}{cll}
  \hline\noalign{\smallskip}
S.No. &  Parameter$^*$       & Adopted values      \\
\hline
	&   Ring                     &               \\
\hline
1 & $R_{\rm in,r}$         &    0.02 pc - 0.8 pc       \\
2 & $\tau_{9.7, \rm r}$  &  0.1-10 \\
3 & $N_{\rm clumps}$     &   $\sim$ 1800, 3000 \\
4 & $R_{\rm out,r}$        &   $R_{\rm in, r}$ + Width \\
\hline\noalign{\smallskip}
	&    Torus               &              \\
\hline
1  & $R_{\rm in,t}$         &   0.03-0.9 pc \\
2  & $\tau_{9.7, \rm t}$     &   10                           \\
3  & $N_{\rm clump}$ & $\sim 8000$   \\
4  & $\sigma$           & 24$^{\circ}$-30$^{\circ}$ \\

\hline\noalign{\smallskip}
	&    Polar Cone               &              \\
\hline
1  & $R_{\rm in,p}$         &   0.07-0.2 pc \\
2  & $\tau_{9.7, \rm p}$     &   10                           \\
3  & $N_{\rm clump}$ & $\sim$ 4000  \\
4  & $\sigma_p$           & 24$^{\circ}$-30$^{\circ}$ \\
\hline
 & Line of sight & \\
 \hline
 1 &  \textit{i} & 0 - 90$^{\circ}$ \\
 \hline

\end{tabular}
\\
$^*$ The other parameters are fixed at p=1, q=0,  F$_{\rm clump}$=0.97 and grain sizes in ring, torus and polar cone are fixed. \\
\end{center}
\end{table}
The dust geometry of NGC 1365 is studied using both RAT and polar models. 
The model SEDs are compared with the observed data from FUV up to a wavelength of 18.3 $\mu$\rm{m} with primary focus on the NIR part of the SED. \\
The parameters are estimated using the goodness of fit test by $\chi^2$ function defined by\\
$\chi^{2} = \frac{1}{(n-p)}\sum\limits_{i=1}^{n} (\frac{\rm O_{i}- \rm M_{i}}{ \rm \Delta_{i}} )^{2}$ \\
where, $\rm O_{i}$ and $\rm M_{i}$ are the observed data and model data for i$^{th}$ photometry point, $\rm \Delta_{i}$ is the observational error, $n$ is the number of observations and $p$ is the number of free parameters.
The parameters like half opening angle, inclination angle, inner radius, and optical depth are varied to get the best fit. \\
Firstly, in order to study the effect of model parameters on the resultant AGN SED, we vary the parameters  $R_{\rm in,r}$ and \textit{i} in the range given in Table \ref{tr1}. The rest of the parameters like N$_{\rm{clumps}}$ in the ring, width, and height of the ring are fixed at 1800, 0.04 pc and 0.04 pc respectively with keeping other parameters at its best fit value i.e. inner radius of the torus at 0.5 pc, half opening angle 26$^\circ$ and optical depth at 10. 
However, none of the models could account for the NIR and MIR features simultaneously. As the number of clumps along the LOS is also a crucial parameter to decide the amount of obscuration, its value for the ring is varied from 1800 to 3000 (around 1.5 times the  initial value). This resulted in a considerably high minimum $\chi^2_{\rm red}$ value of 3.43 and poor fit to the observed SED at MIR wavelengths. The same procedure is followed for a smooth RAT model. We found the clumpy RAT model gives a better fit than the smooth one.
The best fit parameters for clumpy RAT model are found to be
  $\sigma$ = 26$^{\circ}$, $\tau_{9.7, \rm r}$=10, $\tau_{9.7, \rm t}$=10, $R_{\rm in, \rm r}$ = 0.4 pc, $R_{\rm in, \rm t}$=0.5pc, \textit{i} = 69$^{\circ}$, N$_{\rm clumps}$=3000 for L$_{\rm bol}$ = 2.6 $\times$ 10$^{10}$ L$_{\odot}$ (see Figure \ref{fig:fig166}). Now we introduce a polar cone in the RAT model with parameters such as, half opening angle $\sigma_p$, optical depth $\tau_{9.7, \rm p}$, $R_{\rm in, \rm p}$ and $R_{\rm out, p}$. 
   The best fit parameters for the polar model are
  $\sigma_p$ = 30$^{\circ}$, $R_{\rm in, \rm r}$ = 0.03 pc, $R_{\rm in, \rm t}$ = 0.1 pc, $R_{\rm in, \rm p}$ = 0.08 pc, $i$ = 72$^{\circ}$ for L$_{\rm bol}$ = 5 $\times$ 10$^{9}$ L$_{\odot}$ for the size distribution of dust in the graphite ring with 0.1 - 0.5 $\mu$\rm{m}. Figure \ref{fig:fig67} shows the best fit SED for the polar model with $\chi^2_{\rm red}$ value at 1.31. We also found the model with $\chi^2$ of 1.25 after removing the silicate feature for the graphite size distribution of 0.1-1 $\mu$\rm{m} as shown in Fig. \ref{fig:fig67}, but the model is insignificant without the silicate feature.
The comparison between the torus, smooth RAT, clumpy RAT and polar models are presented in Figure \ref{fig:fig511}, where the polar model gives the best-fit SEDs in the entire band based on the minimum $\chi^2$ criterion. The `torus model' and `smooth RAT model' considered here are from \cite{subhashree2021}.

\subsection{Statistical Analysis}
We have used $\chi^2$ for the goodness of fit. We would like to emphasize that optimization of parameters by minimizing $\chi^2$ has not been done in
the present work. We have used $\chi^2$ as a means of comparing the different models. Here, model SEDs have been computed for different combinations of parameters. In order to verify our results for the best parameters obtained from $\chi^2$ analysis, we have performed the R$^2$ analysis for these parameters to study the correlation between the model and the
observed SEDs. To compare the model data and the observed data, we have used Spearman and Pearson correlation analysis where some parameters like size of the clump, half opening angle and $\tau_{9.7}$ are varied. The details of statistical analysis are provided in Table \ref{tab1}. From this table, clumpy RAT model has the highest correlation coefficient with p < 0.05 with the highest $R^2$ coefficient, though all the models give a satisfactory fit to the observed data. But the $\chi^2$ value in the clumpy RAT model is $>$ 2. Hence, the most suitable model would be the polar wind model which is in agreement with $\chi^2$ analysis.
\begin{table}
\begin{center}
\caption{Statistical details of all the model: p is the null-hypothesis probability.} 
\begin{tabular}{llllll}
\hline 
Model  & Pearson coefficient,p & Spearman coefficient, p & R$^2$  \\
\hline 
TO & 0.67,0.05 & 0.73, 0.02 & 0.45 \\
\hline
Clumpy RAT &0.73, 0.02 & 0.72, 0.003 &0.54 \\
\hline
Smooth RAT & 0.72, 0.03 & 0.81, 0.007 & 0.52 \\
\hline
Polar wind & 0.69, 0.04 & 0.82, 0.007 & 0.47 \\  
\hline
\label{tab1}
\end{tabular} 
\end{center}
\end{table}
\section{discussion}
\begin{figure}
\centering
	\includegraphics[width=8cm]{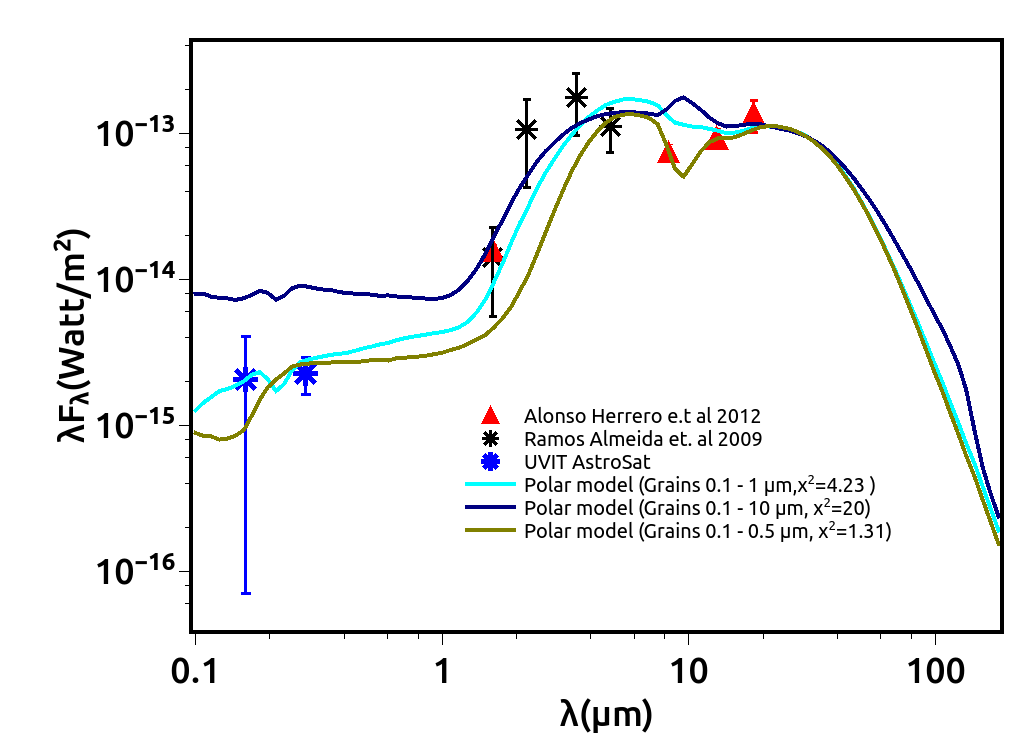}
    \caption{The polar models with different grain sizes are plotted with observed data accounting silicate feature.}
    \label{fig:A2}
\end{figure}

\begin{figure}
\centering
	\includegraphics[width=8cm]{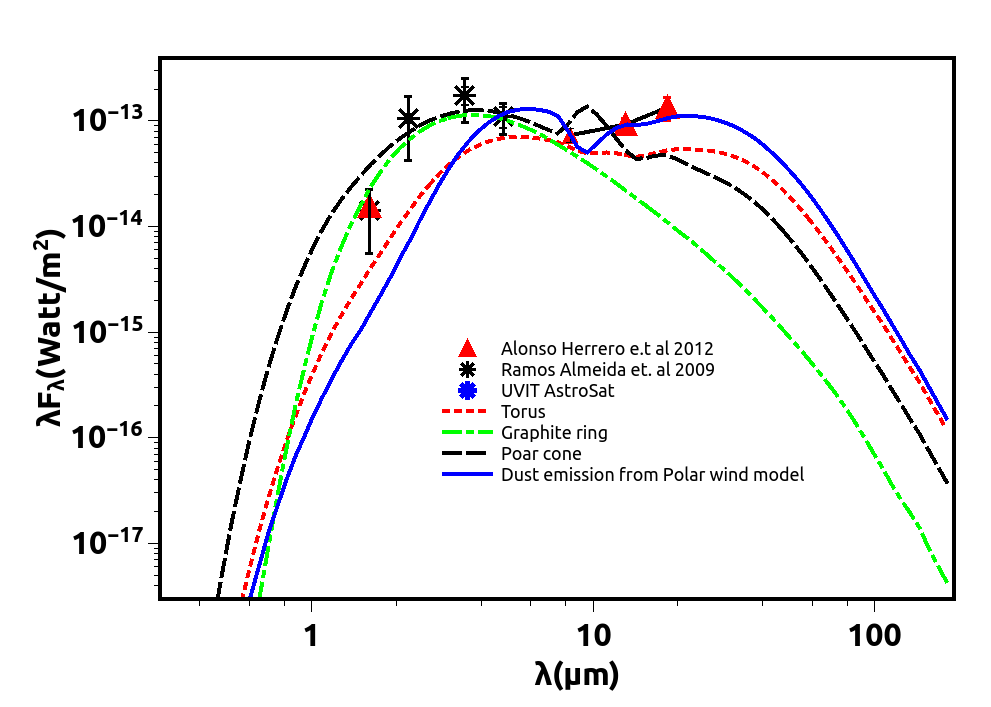}
    \caption{The contributions to the dust components of polar wind model from torus, ring and polar components.}
    \label{fig:fign2}
\end{figure}

The primary motive of this work is to derive the location and characteristics of hot dust in the nuclear region of NGC 1365. The size distribution of graphite grains in the ring is larger than in the \cite{gracia2022} model. However, the clumpy RAT model (two-phase medium of graphite dust grains) and polar wind model used in this work are acceptable models which agree with the $\Delta\chi^2_{\rm red}$ limit of \cite{gracia2022}. Hence the results from polar wind model in the present work are aligned with \cite{gracia2022}, where the authors suggested this model to be the best fit after examining various models.
As fitting the NIR SED is crucial due to the hot dust contribution to this band, we have checked every set of parameters that can provide a better fit to the NIR data.  This work is actually a continuation of \cite{subhashree2021} where authors performed AGN SED fitting for NGC 4151. In the present work, we obtain a minimum $\chi^2_{\rm red}$ value of 1.31 using a realistic model with a dust size distribution of 0.1 - 0.5 $\mu$m in the graphite ring. A larger grain size distribution of 0.1-10 $\mu$m for graphite grains as in \citet{gomar2023} gives a poor fit with $\chi^2_{\rm red}$ = 20 as shown in Figure \ref{fig:A2}. This is due to an increase in the scattered UV intensities from large graphite grains, as the albedo increases. Even a moderately large size distribution in the graphite ring of, 0.1 - 1 $\mu$\rm{m} as in \citet{gracia2022} gives a $\chi^2_{red}$ value of 4.21 which once again represents a poor fit (see Figure \ref{fig:A2}). That model is able to explain the NIR part of the SED since it can match the H and K band intensities, but unable to explain the silicate feature. Hence we conclude that the grain size distribution is a crucial parameter in the fitting procedure when the UV data is also considered. The model SED shown in Figure \ref{fig:fig67} for $R_{\rm{in,r}}$ = 0.03 pc and $R_{\rm{in,t}}$ = 0.1 pc is well fitted with the observed SED in the IR, NUV and FUV wavelength bands. The compact nuclear ring and torus at 0.03 pc and 0.1 pc respectively lie well inside the molecular torus, which has a size of 26 pc based on CO(3-2) observations \citep{combesetal2019}. Our model predicts the presence of two distinct dust populations in agreement with reverberation mapping observations of \cite{Lyu_2021}. The minimum radius of the polar cone, 0.08 pc places it between the graphite ring and the torus. 
Our best fit parameters are compared with those from the literature in Table \ref{tab:tab0}. The inner radius of the sublimation zone is calculated by \cite{fritz2006} and \cite{selfcon} to be 0.17 pc and 0.24$^{+0.21}_{-0.18}$ pc respectively. These radii are in agreement with the results from the current work, where the inner radius of torus R$_{\rm in,t}$ is 0.1 pc. Further, we focused on hot dust in the regime near the central nucleus. Here, the situation demands an additional ring structure of large grains in the inner sublimation region at inner radius R$_{\rm in, r}$ = 0.03 pc. The changes from 0.1 pc to 0.03 pc (from MIR to NIR) could be due to the luminosity which changes the environment from smooth/clumpy to two-phase. This result is also supported by UV emission from the disk where dust can sublimate at 0.03 pc, 0.08 pc and 0.1 pc. The best fit radius for the torus 0.1 pc from our work is in agreement with the values proposed by different models. The polar model consists of three components torus, ring and polar. We found that a single component cannot explain the total emission. Hence we expect the polar wind component as the major contributor to the emission as shown in Figure \ref{fig:fign2}.  \\
\begin{table*}
\begin{center}
\caption[]{The list of derived parameters for NGC 1365 from the literature presented along with the best fit values from this work.  For reference , 1: \cite{fritz2006}, 
2:\cite{Ramos_Almeida_2011}, 3: \cite{hererroetal2012}, 
4: \cite{selfcon}, 5: \cite{gracia2022}, 6.\cite{herrero2021}. Column 1 shows the inclination angle $i$, column 2 represents the half opening angle $\sigma$, column 3 and 4 represents the dust distribution parameters (p, q), column 5 and 6 are the optical depth of the dust ($\tau_{9.7}$,$\tau_v$) which are normalized at 9.7 $\mu$m and visible wavelength respectively. Column 7 shows other parameters, not common to all the models, where $\gamma$ is radial extent of the torus, A$^{LOS}_v$ is the optical extinction at line of sight, N$_{0}$ is the number of clouds along the equatorial ray, $\eta$ is filling factor for the torus, $\tau_{v,cl}$ is the cloud optical depth, $\tau_{v,mid}$ is the optical depth of the disk mid-plane, a is the size of grain and $a_w$,$h$,$f_w$ are the wind parameters used in \cite{2017ApJ...838L..20H}.} 
\begin{tabular}{ccccccp{4cm}ll}
\hline
$i$(degrees)                 & $\sigma$(degrees)       & p                    & q & $\tau_{9.7}$ & $\tau_{v}$ & Other parameters & Model \\ 
\hline
- & 70 & 0 & 0 & 3 & - & R$_{min}$=0.17 pc, R$_{max}$=52 pc, A$_v$=64, N$_{H}$=2.59 $\times$ 10$^{23}$ cm$^{-2}$, M$_{dust}$=7.5 $\times$ 10$^4$M$\odot$ & [1](Smooth torus model) \\
\hline
27$^{+19}_{-16}$ & 35$^{+14}_{-10}$ & - & 1.1$^{+0.8}_{-0.7}$ &- & 86$^{+36}_{-38}$ &A$_v^{LOS}$<110, $\gamma$=18$\pm$7, N$_{0}$=7$\pm$4   &[2](Clumpy torus model)\\
\hline

- &        36$^{+14}_{-6}$         &          -          & - & - & - & R$_{out}$=5$^{+0.5}_{-1}$, L$_{bol}$ = 2.6$\pm$0.5 $\times$ 10$^{43}$ ergs sec$^{-1}$ & [3](Clumpy torus model)\\
\hline
33$^{+10}_{-14}$ &        -          &      -              &- &- & 14$^{+31}_{-9}$ ($\tau_{v,cl}$) &R$_{\rm in}$=0.24$^{+0.21}_{-0.15}$, $\eta$=8$^{+67}_{-7}$, $\tau_{v,mid}$= 470$^{+530}_{-430}$, Intrinsic L$_{AGN}$(LogL$\odot$)= 10.23$^{+0.05}_{-0.04}$ & [4](Two phase model)\\
\hline
52$^{+7.6}_{-2.6}$ & $>$12.6 & $<$--3.0 & - & -&- & $a_w$=-1.58$^{-0.06}_{-0.25}$, $h$= $<$0.1, $f_w$= $>$0.6, $N_0$= $<$6.8, $\theta$>42.6 &[5](Clumpy disk$+$wind model)\\
\hline
8-40, 53 & & & & & & i=$>$48$^\circ$ from ALMA imaging & [6]Clumpy, Xclumpy\\

\hline
72 & 26 & 1 & 0 & 20 &- & $R_{\rm in,r}$ = 0.03 pc, $R_{\rm in,t}$= 0.1 pc, L$_{AGN}$ = 5 $\times$ 10$^{9}$ L$_{\odot}$, N$_{\rm clumps}$ = 1800 & This work \\
\hline
\label{tab:tab0}
\end{tabular} 
\end{center}
\end{table*} 

\begin{figure}
\centering
	\includegraphics[width=8cm]{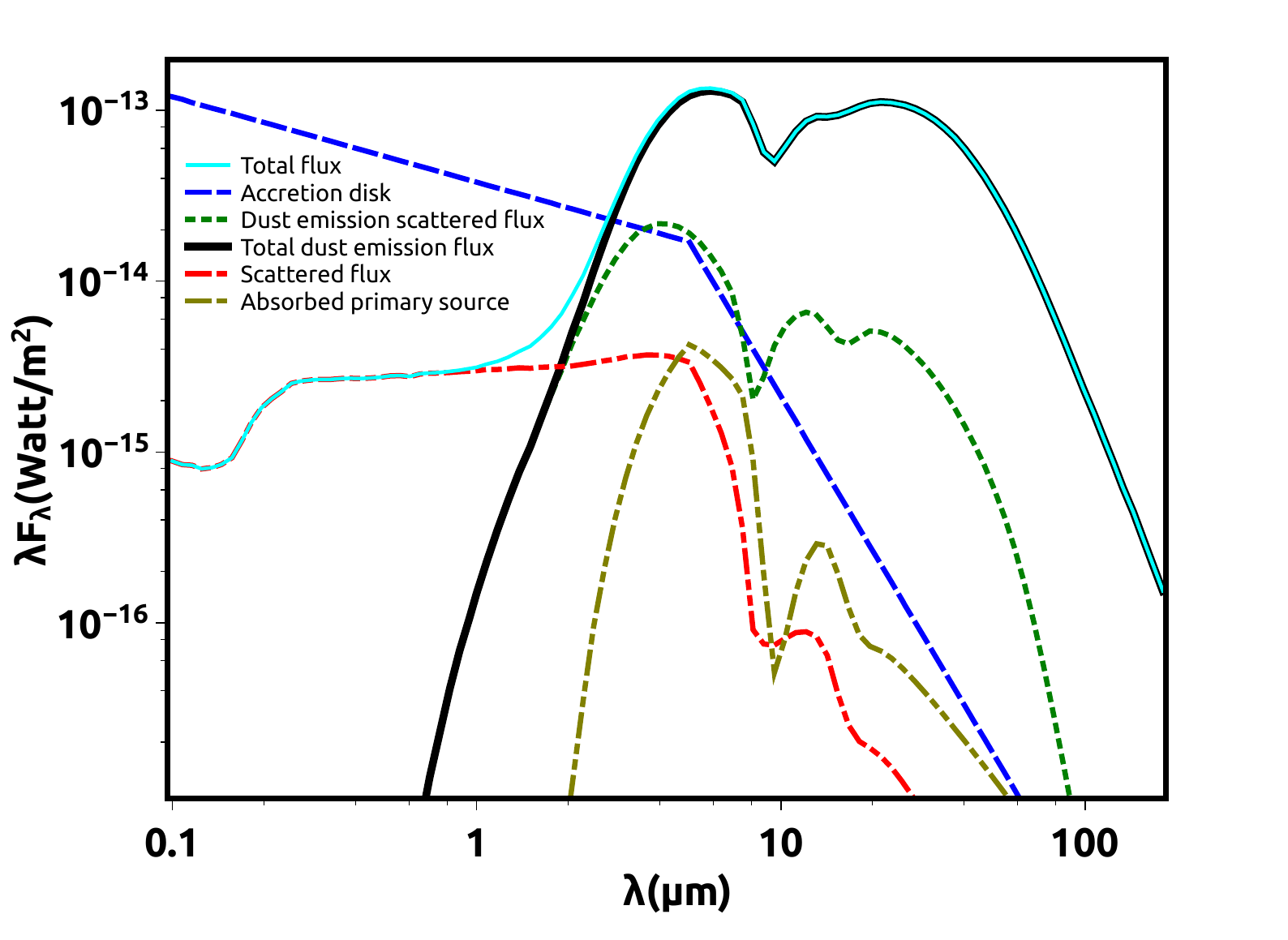}
\caption{The contribution from the accretion disk, the dust geometry and the scattered emission to the SED for polar model, respectively, at its best fit parameters. The scale on X- and Y-axis is logarithmic.}
\label{fig:fig26}
\end{figure}

\begin{figure}
\centering
	\includegraphics[width=8cm]{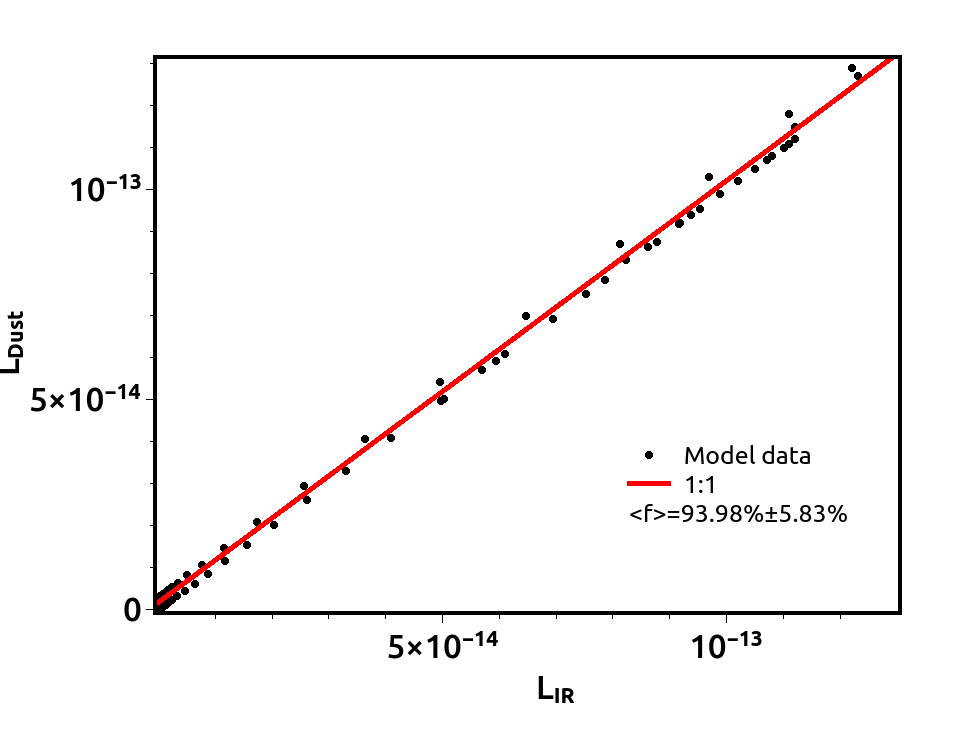}
\caption{Relation between total IR luminosity (L$_{IR}$ ) and the
luminosity of dust in the AGN (L$_{Dust}$) derived for the best fit model. The median
luminosity fraction of the AGN f$_{AGN}$ = 93.98\%$\pm$5.83\%.}
   	\label{fig:fig34}
\end{figure}

\begin{figure}
\centering
	\includegraphics[width=8cm]{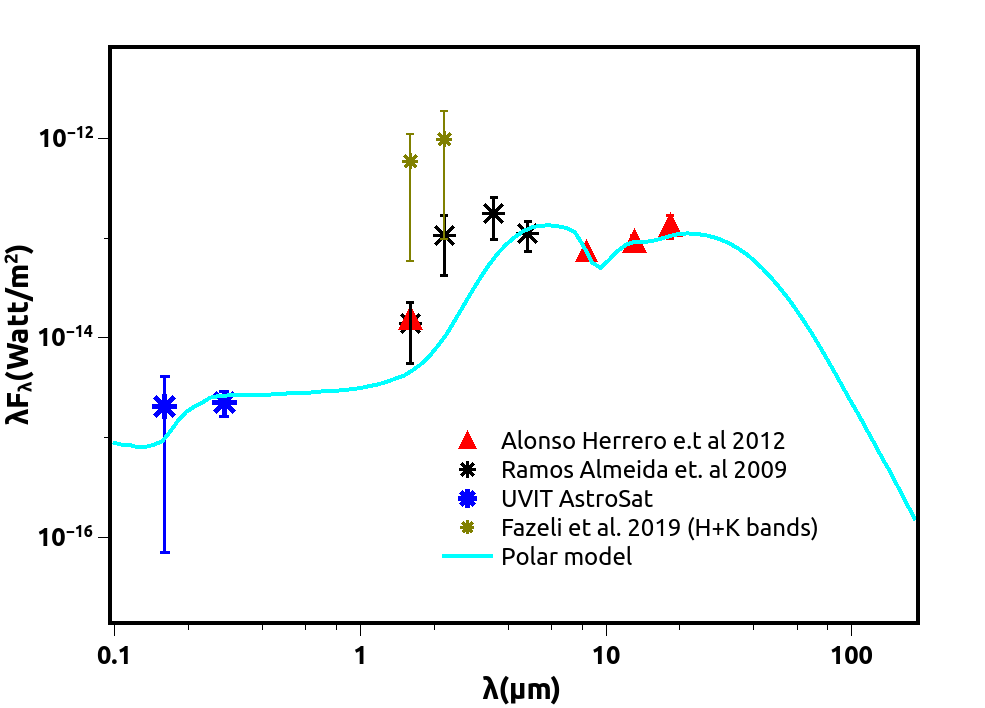}
\caption{The stellar contaminated observation in H$+$K band compared with the polar model.}
\label{fig:fig52}
\end{figure}

All the components contributing to the optical/IR emission are displayed in 
Figure \ref{fig:fig26}. It shows the absorbed UV radiation is re-emitted in the NIR and MIR bands. In the UV/NIR part of the SED, the accretion disk also contributes together with the emission and scattering from hot dust in the ring/torus. However, our best fit model is unable to explain the NIR excess emission in the central nuclear region as being entirely due to hot dust grains. The mismatch in the H band is probably due to contamination from stellar emission as this AGN is known to host star formation in its circumnuclear regions. 
The emission in the NIR band is a mixture of hot dust emission and stellar emission. The resolution of these emissions from the total IR SED is beyond the scope of this work. The visibility of hot dust in type 2 AGN is less than that in type 1 AGN which is reflected by the steepness of SED in the NIR region. However, in spite of the fact that this AGN is of type 
1.8/2, the hot dust is visible because of the clumpiness of the medium.
The one to one relation between the IR luminosity (L$_{\rm{IR}}$) and dust luminosity (L$_{\rm{Dust}}$) is presented in Figure \ref{fig:fig34} for the best fit model. It is evident from the figure that 93.98\%$\pm$5.83\% of total IR luminosity L$_{\rm{IR}}$ is due to the dust in the AGN. The excess IR luminosity for $\lambda$ $<$ 2$\mu$\rm{m} can be attributed to stellar contamination. \cite{fazeli2019} observed strong Pa$\alpha$ emission at 1.8 $\mu$\rm{m} and several other stellar emission lines in H and K bands using ESO's SINFONI instrument (see Figure \ref{fig:fig52}).\ Hence accounting for the stellar  contributions in the polar model could significantly improve the SED fit in NIR. Adding 10\% error as stellar contamination to the H and K bands in the observed SED improves the polar model fit to the SED with a $\chi^2_{\rm red}$ value of 1.22.


The most noteworthy feature in Seyfert AGN in MIR band is the silicate feature at 9.7 $\mu$\rm{m}. In most typical type 1 and type 2 AGN, the silicate feature appears as emission and absorption peaks respectively.
There is no confirmation about the 9.7 $\mu$\rm{m} silicate feature in NGC 1365 based on GEMINI/T-ReCS observations \citep{hererroetal2012, herrero2020, herrero2021}.  
All the models with silicate features are shown in Figure \ref{fig:fig511}, where the best fit polar model produced an absorption silicate feature because dust grains in the polar wind absorb the UV radiation from the accretion disk. But if we remove the polar wind component, we see a mild absorption/flat silicate feature in TO and RAT models. The standard ISM silicate dust in the torus and large grains in the polar cone could cause the IR absorption at 9.7 $\mu$m as seen in Figure \ref{fig:fig511}. Our polar model with a silicate absorption feature resulted in a $\chi^2_{\rm red}$ of 1.31 when we include the 9.7 $\mu$m data, but with a mismatch in the NIR. However, when the 9.7$\mu$\rm{m} data is not included in the fit, we found $\chi^2_{\rm red}$ of 1.25 with a flat MIR in the SED (see Fig. \ref{fig:fig67}).  The MIR emission is considered to be due to UV light being reprocessed by the hot dust surrounding the AGN. It is known that the dust distribution, composition and UV luminosity decide the nature of the silicate feature (see Figure \ref{fig:fig511}) and it is not a simple function of inclination angle \citep{sturm2005, stalevski2012}. As UV emission is not variable, the changes in the MIR silicate feature in our model is due to different dust grain size distributions.\\

\subsection{Location of Hot Dust}
\begin{figure}
\centering
	\includegraphics[width=8cm]{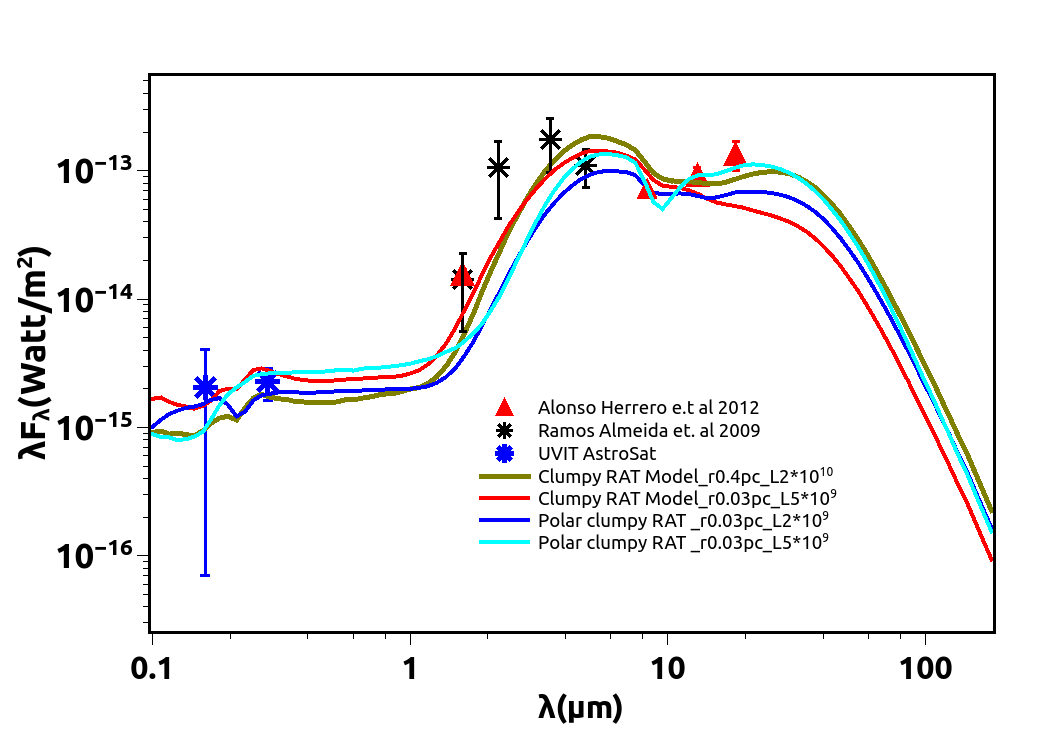} 	
\caption{Comparison between all the models with varying different source luminosities as input.}
\label{fig:A1}
\end{figure}

The best fit inner radii of the ring and torus are 0.03 pc and 0.1 pc respectively for the polar model while they are 0.4 pc and 0.5 pc respectively for the clumpy RAT model. 
In order to find the location of hot dust in this AGN with this model, we further performed more simulations with the inner radius set to 0.02-0.06 pc only to understand the flat nature of NIR SED. The parameters like optical depth, number of clumps, and inclination angle are varied. The best fit is obtained at 
$\sigma$ = 26$^{\circ}$, $\tau_{9.7, \rm r}$=0.1, $\tau_{9.7, \rm t}$=5, $R_{\rm in, \rm r}$ = 0.03 pc, $R_{\rm in, \rm t}$=0.4 pc, \textit{i} = 69$^{\circ}$, N$_{clumps}$ = 3000, $p$ = 1 and $q$ = 0 with poor reduced $\chi^2$ > 4. Hence, we conclude that the clumpy RAT model does not give a good fit to the observed SED of NGC 1365 and does not explain the observed inner radius even if the input luminosity is varied as shown in Figure \ref{fig:A1}. 
\begin{figure}
\centering
	\includegraphics[width=8cm]{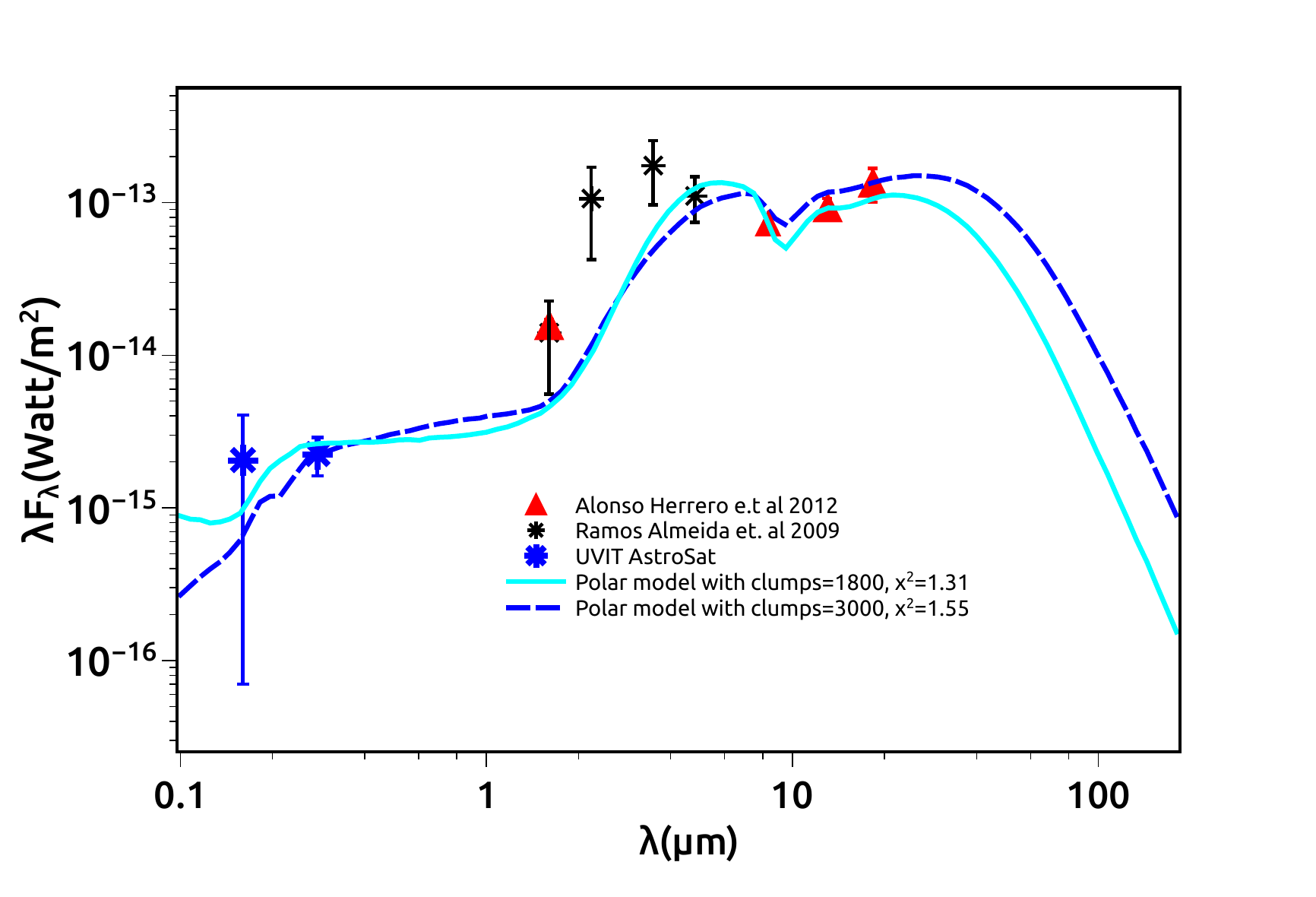} 	
\caption{Comparison between the best-fit polar Model (in cyan color) with $N_{\rm clumps}$ = 1800 and the model with same parameters with $N_{\rm clumps}$=3000
 in blue color.}
\label{fig:fig16}
\end{figure}
However, incorporating the polar cone into the RAT model significantly improves the fit, with 
hot dust at inner radius of the ring, 0.03 pc and warm dust at inner radius of the torus, 0.1 pc. Also, the best fit inner radius of 0.03 pc is able to explain the L and M band fluxes, that are dominated by AGN emission. Hence including pure graphite dust in the dust sublimation model proves to be an established method \citep{subhashree2021, gracia2022} to reproduce the nuclear NIR emission of AGN. 
The hot dust near the nucleus \citep{dexter2020} can be detected in the face-on orientation when NGC 1365 is considered as a Seyfert 1 AGN or in the edge-on direction as a Seyfert 2 AGN due to the clumpy medium.
However, in this case, the observed SED seems to be of type 2 AGN as the SED steepens at $\lambda$ $<$ 2 $\mu$\rm{m}. Thus, NGC 1365 should contain a clumpy dust medium so as to detect the hot dust at 0.03 pc. In addition, the number of clumps in the graphite ring is found to be an important parameter as shown in Figure \ref{fig:fig16}. The intrinsic UV luminosity of this AGN observed by the UVIT is $\sim$76 times lower than the source luminosity used in the polar model. We find the graphite sublimation radius to be $\sim$ 0.003 pc using UVIT source luminosity which puts it inside the BLR region. In fact, the luminosity affects the inner radius in the sublimation zone. The source luminosity in the clumpy RAT model is 1.23 times higher than in the polar model. Thus, input  luminosity is a vital factor for the location of hot dust. 
An illustration based on the results of this model is shown in Figure \ref{fig:fig17} where our model derived radius of 0.03 pc is compared with the observed inner radius from the \cite{dexter2020}. We find that torus inner radius at 0.1 pc is well in agreement with \cite{tristam2011}.  The clumpy dust clouds in the torus are unable to account for the observed NIR emission with standard ISM dust whereas the clumpy graphite dust clouds in the polar model overcome this by modifying the dust distribution and radiative transfer effects \citep{Hatzim2015} in the NIR. Although the PAH emission at 3.5 $\mu$\rm{m} wavelength is known as a good starburst indicator, the emission at this particular wavelength may contain both the hot dust and stellar components in our case as suggested by \cite{Woo2012}. The hot dust modelling in this starburst AGN has not been widely investigated in the past since it was difficult to explain such a scenario. In summary, the excellent resolution of UVIT and highly sophisticated IR observations allowed us to fit the observed SED with hot dust at 0.03 pc and torus at 0.1 pc from the center in a sublimation zone consisting of two regions with dust temperatures of 1216 K and 914 K respectively.
\begin{figure}
\centering
	\includegraphics[width=8.5cm]{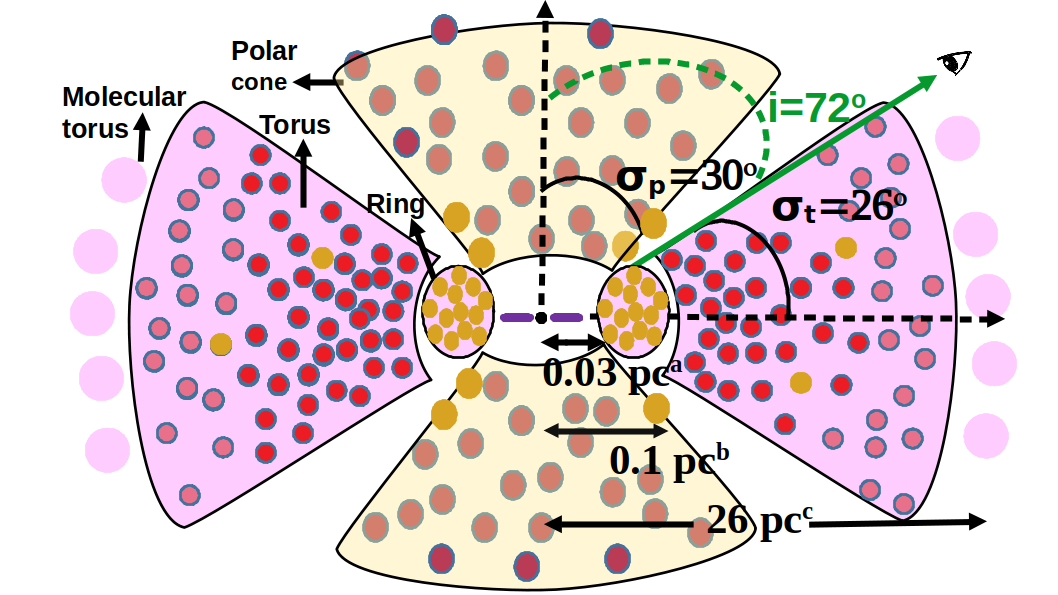} 	
\caption[]{A schematic illustration of polar model included with literature based inner radii of dust.
Reference: {\bf \lowercase{A}}: \cite{dexter2020} and our best fit inner radius of the ring, {\bf \lowercase{B}}: Best fit inner radius of the torus and \cite{tristam2011}, {\bf \lowercase{C}}: \cite{combesetal2019}.}
\label{fig:fig17}
\end{figure}

\section{Conclusions}
In this paper, we have investigated the location of hot dust in a sublimation zone in the innermost regime of the dusty torus of AGN NGC 1365 using the {\sc skirt} code based polar model with a two phase smooth and clumpy medium. We have constructed the UV-IR SED of this AGN by incorporating high resolution UVIT observations in addition to archival IR observations.
We have demonstrated that a two phase medium with a dust geometry of the sublimation zone, the torus and polar cone defined by $R_{\rm in,r}$ = 0.03 pc, $R_{\rm in,t}$= 0.1 pc, $R_{\rm in,p}$ = 0.08 pc, $\tau_{9.7, \rm r}$ = 10, $\tau_{9.7, \rm t}$ = 10, $\sigma_t$ = 26$^{\circ}$, $\sigma_p$ = 30$^{\circ}$, i=72$^{\circ}$ and N$_{\rm clumps}$ = 1800 gives a reasonable fit to the observed UV-IR SED of NGC 1365 with a minimum $\chi^2_{\rm red}$ value of 1.31.
The location of graphite dust is found to be at 0.03 pc from the central source which is in agreement with observations \citep{dexter2020}. 
The hot sublimation zone has temperatures of 1216 K and 914 K for different dust grain populations and can reproduce the observed absorption silicate feature. Since UV scattering by dust, which is a non-negligible component of the UV SED of AGN, depends on the grain size distribution, we find that fitting the UV-IR SED helps in constraining the size of the grains. We find the detection of hot dust emission in the sublimation zone of this Seyfert 1.8 AGN to be due to the clumpiness of the medium. Our polar model deviates slightly from the observed SED in the NIR H and K bands that are known to be contaminated by stellar emission. The fit can be improved further if this stellar contamination can be accurately determined in the unresolved nuclear region of this AGN.
\section*{Acknowledgements}
We thank G.C. Dewangan for allowing us to use his \textit{AstroSat} observations as part of the proposal ID A02006 (PI:Gulabd).
One of the authors (PS) acknowledges Manipal Centre for 
 Natural Sciences, MAHE for facilities and support.


\section{Data Availability}
The UVIT data from the three observations used in this paper are publicly available at the
AstroSat data archive \url{https://astrobrowse.issdc.gov.in/astro_archive/archive/Home.jsp} maintained by the ISSDC. The compiled IR  observed data underlying this article can be retrieved from \cite{almeida_2009, hererroetal2012}. The original observed data are available in SIMBAD via link \url{http://simbad.u-strasbg.fr/simbad/sim-ref?querymethod=bib&simbo=on&submit=submit+bibcode&bibcode=2012MNRAS.425..311A} and \url{http://simbad.u-strasbg.fr/simbad/sim-ref?querymethod=bib&simbo=on&submit=submit+bibcode&bibcode=2009ApJ...702.1127R}. \\
The model SED data will be shared on a reasonable request to the authors(\url{s.subhashree00@gmail.com}).



\bibliographystyle{mnras}
\bibliography{reference} 

\bsp	
\label{lastpage}
\end{document}